\documentclass{article}[10pt]
\usepackage{amssymb}
\usepackage{amsmath}
\usepackage{epsfig}
\usepackage{hyperref}
\usepackage{multirow,bigdelim}
\usepackage{array}
\usepackage{enumerate}
\usepackage{url}
\usepackage{verbatim}
\usepackage{ctable}

\newcommand{\parab}[1] {\bigskip \noindent \textbf{#1}~}
\newcommand{\param}[1] {\medskip \noindent \textbf{#1}~}

\newcommand{\otoprule} {\midrule[\heavyrulewidth]}

\newfont{\helvetica}{phvr8t at 10pt}
\newfont{\helveticasmall}{phvr8t at 9pt}
\newfont{\helveticabig}{phvr8t at 22pt}
\newfont{\helveticao}{phvro8t at 11pt}

\ifdefined\labelenumiii
  \renewcommand{\labelenumiii}{(\roman{enumiii})}
\else
  \newcommand{\labelenumiii}{(\roman{enumiii})}
\fi

\date{}
\title{Query Expansion Using Term Distribution and Term Association}
\author{Dipasree Pal, Mandar Mitra \\ Indian Statistical Institute, Kolkata 
        \and Kalyankumar Datta \\ Jadavpur University, Kolkata}

\begin{document}
\maketitle

\begin{abstract}
  Good term selection is an important issue for an automatic query
  expansion (AQE) technique. AQE techniques that select expansion terms
  from the target corpus usually do so in one of two ways.
  \emph{Distribution based} term selection compares the distribution of a
  term in the (pseudo) relevant documents with that in the whole corpus /
  random distribution. Two well-known distribution-based methods are based
  on Kullback-Leibler Divergence (KLD)~\cite{carpineto-kld} and
  Bose-Einstein statistics (Bo1)~\cite{amati-thesis}. \emph{Association
    based} term selection, on the other hand, uses information about how a
  candidate term co-occurs with the original query terms. Local Context
  Analysis (LCA)~\cite{lca} and Relevance-based Language Model
  (RM3)~\cite{rm3} are examples of association-based methods. Our goal in
  this study is to investigate how these two classes of methods may be
  combined to improve retrieval effectiveness. We propose the following
  combination-based approach. Candidate expansion terms are first obtained
  using a distribution based method. This set is then refined based on the
  strength of the association of terms with the original query terms. We
  test our methods on 11 TREC collections. The proposed combinations
  generally yield better results than each individual method, as well as
  other state-of-the-art AQE approaches. En route to our primary goal, we
  also propose some modifications to LCA and Bo1 which lead to improved
  performance.
\end{abstract}

\section{Introduction}
Consider a user's query $Q$ and a relevant document $D$ from a document
collection. $Q$ and $D$ may use different vocabulary to refer to the same
concept. Information Retrieval (IR) systems that rely solely on
keyword-matching may not detect a match between $Q$ and $D$, and may
therefore not retrieve $D$ in response to $Q$. This is the well-known
\emph{vocabulary mismatch} problem in IR.

A good retrieval system must solve this problem by bridging the vocabulary
gap that exists between useful documents and the user's query. Query
Expansion (QE) is an important technique that attempts to increase the
likelihood of a match between the query and relevant documents by adding
related terms (called \emph{expansion terms}) to a user's query.

A wide variety of methods for Automatic Query Expansion (AQE) have been
proposed over the last 15--20 years. These methods find related terms from
different sources such as the target corpus, linguistic resources like
Wordnet~\cite{wordnet}, thesauri~\cite{mandala}, ontologies~\cite{onto},
the World Wide Web, Wikipedia~\cite{wiki} and query logs~\cite{ql}. A recent
survey of such techniques can be found in~\cite{qe-survey}. Of all
these techniques, methods that use the target corpus as a source of
expansion terms are among the most widely used because they are simple and
require no additional resources. 

Target-corpus-based AQE techniques can be broadly classified into two
\linebreak groups:
\emph{distribution based} and \emph{association based}. Distribution 
based methods select terms by comparing the distribution of a term in the
(pseudo) relevant documents with its distribution in the whole corpus.
Broadly, such methods select terms that are more likely to occur in the
(pseudo) relevant documents than in a document chosen randomly from the
entire corpus.
On the other hand, association based methods select expansion terms on the
basis of their association (or co-occurrence) with all query terms. A term
that tends to co-occur with all / many of the query terms is regarded as a
good expansion term.

While a number of distribution / association-based QE techniques have been
shown to be effective on average (i.e.\ when their overall performance
across a large set of queries is measured), the impact of different QE
techniques on individual queries can vary greatly.

\begin{table}[h]
\scriptsize
  \centering
  \begin{tabular}{ccccc}\toprule
             & Baseline & Assoc.\ based & Distr.\ based    \\\otoprule
   MAP       & 0.218    & 0.250 (+14.8) & 0.257 (+ 18.0\%) \\
   Better on & --       & 81 queries    & 91 queries       \\\bottomrule
 \end{tabular}
 \caption{{\scriptsize Potential improvement obtainable in principle by judiciously choosing QE techniques}}
 \label{tab:variability}
\end{table}
\normalsize

Table~\ref{tab:variability} shows the Mean Average Precision (MAP) scores
for three retrieval methods on TREC queries 301--450 (for more details,
please see Section~\ref{sec:exp-set}): a baseline strategy that uses original,
unexpanded queries, and representative distribution-based~\cite{carpineto-kld}
and association-based~\cite{lca} QE methods. The QE methods are superior to
the baseline on average, but they result in decreased performance for a
number of queries. Further, while the overall performance figures for
these two QE methods are comparable, each of these methods outperforms the
other on about half the queries used in this experiment.

As these two methods work in different ways, our hypothesis is that if we
combine these two methods by considering both distribution information and
association information, we should be able to improve overall performance.
In this study, therefore, we investigate the possibility of improving
retrieval effectiveness by combining association- and distribution-based
QE approaches. We first select two well-known, representative method from
each category, viz.\ LCA~\cite{lca}, RM3~\cite{rm3} (association-based) and
KLD~\cite{carpineto-kld}, Bo1~\cite{amati-thesis} (distribution-based). 
Next, we introduce some simple modifications in the basic formulae of some
of these methods in order to improve their performance. We verify that
these modifications indeed result in better retrieval effectiveness. 
Finally, the two approaches are combined as follows: we select a relatively
large number of candidate expansion terms using the distribution based
method. Some of these are filtered out using information from the
association based method. The refined set is finally used for query
expansion.

We test our combined method on eleven TREC collections. Our proposed method
yields significant improvements on all collections over a baseline that
uses the original, unexpanded queries. More importantly, the combined
methods yield improvements over the individual AQE methods for most of the  
collections. 

In summary, this study makes the following contributions.
\begin{itemize}
\item It proposes refinements for some well-known QE methods.
\item It demonstrates that a combination of distribution based and
  association based methods outperforms the individual methods as well as
  state-of-the-art QE methods, such as the approaches proposed in
  \cite{rm3,amati-thesis}.
\end{itemize}

In the next section, we discuss the relationship between this study and
related work. Section~\ref{sec:method} briefly reviews the existing AQE
methods that are used in this study, our modifications of these methods, 
as well as the proposed method for combining AQE techniques. Section
\ref{sec:exp-set} describes the experimental setting that we used.  
Results comparing the proposed methods with existing ones are presented in
Section~\ref{sec:exp-res}. Finally, Section~\ref{sec:conclusion} summarizes
some related issues that need to be studied in future work.

\section{Related work}
Early work on automatic query expansion dates back to the 1960s. Rocchio's
relevance feedback method~\cite{rocchio71} is still used in its original
and modified forms for AQE. The availability of the TREC collections, and
the widespread success of AQE on these collections stimulated further
research in this area. Carpineto and Romano~\cite{qe-survey} provide a
recent and comprehensive survey of AQE techniques. We focus here on some
important AQE techniques that are either distribution- or association-based.

\param{Association-based QE techniques.} Early work on association-based
AQE includes ``concept-based'' QE~\cite{qiu-frei} and
\emph{phrasefinder}~\cite{phrasefinder}. Both methods make use of term
co-occurrence information extracted from a corpus. Local context analysis
(LCA)~\cite{lca, lgca} is another well-known method that also selects
expansion terms based on whether they have a high degree of co-occurrence
with all query terms. However, in LCA, co-occurrence information is
obtained from a set of top-ranked documents retrieved in response to the
original query, rather than the whole target corpus. Relevance-based
language models~\cite{rblm} constitute another, more recent, co-occurrence
based approach. This method is based on the Language Modeling framework.
The query and relevant documents are all assumed to be generated from an
underlying \emph{relevance model}. This model is estimated based on (only)
the pseudo relevant documents for a particular query. This approach was
subsequently refined by AbdulJaleel et al.~\cite{rm3}. The refinement,
called RM3, incorporates the original query when estimating the relevance
model. According to a comparative study by Lv et
al.~\cite{comp-prf,roch-prox}, RM3 is the most effective and robust among a
number of state-of-the-art AQE methods. RM3 is frequently used as a
baseline against which several recent QE methods have been
compared~\cite{roch-prox,posrel,concept-weight,good-term,inter-passage}.

\param{Distribution-based QE techniques.} As early as 1978,
Doszkocs~\cite{doszkocs} proposed the interactive use of an associative  
dictionary that was constructed based on a comparative analysis of term
distributions. Also well known is Robertson's analysis of term selection
for query expansion~\cite{robertson90}. More recently, Carpineto et
al.~\cite{carpineto-kld} proposed an effective QE method based on
information theoretic principles. This method uses the Kullback-Leibler
divergence (KLD) between the probability distributions of terms in the
relevant (or pseudo-relevant) documents and in the complete corpus.

Amati~\cite{amati-thesis} proposes a new distribution based method which
uses Bose-Einstein statistics. This method also calculates the divergence
between the distribution of terms in the pseudo relevant document set and a
random distribution.

Efforts have also been made to combine AQE methods in various ways to
improve retrieval effectiveness. Carpineto et al.~\cite{carpineto-mult-rank} 
combined the scoring functions of a number of methods, all of them
distribution-based, to obtain improvements. In contrast, we combine a
distribution-based method with an association-based method (based on our
belief that these two classes of methods offer different advantages). Also,
rather than combining scores, we use one method to refine the set of terms
selected by the other.\footnote{Of course, this can also, strictly
  speaking, be regarded as a combination where one component is very highly
  weighted.} Our approach is somewhat similar in spirit to a method
proposed by Cao et al.~\cite{sel-good-term}, in which terms selected using
standard pseudo relevant feedback (PRF) are refined using a classifier that
is trained to differentiate between useful and harmful candidate expansion
terms. Our work is most strongly related to that of P{\'e}rez-Ag{\"u}era and 
Araujo~\cite{compcomb}, who also combine co-occurrence-based and
distribution-based methods. The combination is relatively straightforward:
one method is used for term selection and the other for weighting. Word
co-occurrence is measured using the Tanimoto coefficient. Distributional
differences are measured based on KLD or Bose-Einstein statistics. The
methods are tested on a relatively small Spanish dataset. We use the
well-known LCA and RM3 method (instead of Tanimoto coefficient) to quantify
term association. Also, instead of simply using one method for term
selection and the other for weighting, we combine both methods for
selection. Finally, we test our method on a number of large TREC datasets.

\section{Methods}
\label{sec:method}
\subsection{Basic Methods}
\label{sec:basic-methods}
We first review  KLD, Bo1, LCA and RM3, the existing methods that form the
base of our approach.

\subsubsection{Distribution based method I: KLD}
\label{sec:kld}
The approach proposed by Carpineto et al.~\cite{carpineto-kld} is one of
the two distribution based term ranking methods used in this study. In this
method, all terms in the pseudo relevant set are treated as candidate
expansion terms. Let $R$ and $C$ represent the (pseudo) relevant documents
(PRD) and the whole corpus respectively. We use $p_r$ and $p_c$ to denote
the unigram probability distribution of terms in $R$ and $C$ respectively;
$p_r$ and $p_c$ are calculated as shown in Equations~\ref{pr} and
\ref{pc} ($\it{tf}(t,d)$ represents the term frequency of term $t$ in
document $d$).
\begin{equation}\label{pr}
  p_r(t) = \frac{\sum\limits_{d \in R}\it{tf}(t,d)}{\sum\limits_{d \in R}\sum\limits_{t^\prime \in d}\it{tf}(t^\prime,d)}\\
\end{equation}

\begin{equation}\label{pc}
  p_c(t) = \frac{\sum\limits_{d \in C}\it{tf}(t,d)}{\sum\limits_{d \in C}\sum\limits_{t^\prime \in d}\it{tf}(t^\prime,d)}  \\
\end{equation}
The contribution of a term to the divergence between $p_r$ and $p_c$ is
given by Equation~\ref{kld-eq}. Terms for which this contribution is the
largest are selected as expansion terms.
\begin{equation}\label{kld-eq}
  S(t) = p_r(t) * \log{\frac{p_r(t)}{p_c(t)}} \\
\end{equation}
$S(t)$ is also used as the term weight of a candidate expansion term $t$. 

\medskip
In our experiments with KLD (and other methods), we use
Equations~\ref{s-orig} to~\ref{f-score} to merge the original query terms 
with the candidate expansion terms to formulate the final expanded
query. The weights of original query terms are normalized using the maximum
original query term weight (Eqn.~\ref{s-orig}); weights of expansion terms
are similarly normalized (Eqn.~\ref{s-exp}). These weights are simply added
together to obtain the final weight of a term $t$ in the expanded query
(Eqn.~\ref{f-score}). 

\begin{equation}\label{s-orig}
 score_{orig}(t) = \frac{1+\log(\it{tf(t,Q)})}{1 + \max\limits_{t^\prime \in Q} \log(\it{tf(t',Q)})}
\end{equation}
\begin{equation}\label{s-exp}
 score_{exp}(t) = \frac{S(t)}{\max\limits_{t^\prime \in d \in {PRD}} S(t^\prime) }
\end{equation}
\begin{equation}\label{f-score}
 score(t) = score_{orig}(t) + score_{exp}(t) 
\end{equation}

\subsubsection{Distribution based method II: Bo1}
\label{sec:bo1}
The second, more recent, distribution based term ranking model we
considered is Bo1, which is the most effective variant of the Divergence
From Randomness (DFR) term weighting
model~\cite{DBLP:conf/trec/PlachourasHO04,DBLP:conf/trec/MacdonaldHPO05}.
In this model based on Bose-Einstein statistics, the informativeness of a
term $t$ is measured by the divergence between its distribution in the top
ranked documents and a random distribution. Specifically, the score of a
candidate expansion term $t$ is given by
\begin{equation}\label{bo1}
  S(t) = \left(\sum_{d \in \mathit{PRD}} \mathit{tf}(t,d)\right) * \log_2{\left(\frac{1+f_{avg}(t,C)}{f_{avg}(t,C)}\right)} + \log_2{(1+f_{avg}(t,C))} \\
\end{equation}
where
\begin{equation}
  f_{avg}(t,C) = \sum\limits_{d \in C}\mathit{tf}(t,d)/N
\end{equation}
denotes the average term frequency of $t$ in the collection ($N$ is the
number of documents in the collection). As in Section~\ref{sec:kld}, we use
Equations~\ref{s-orig}--\ref{f-score} to merge the original query with the
expansion terms and formulate the new expanded query.

\subsubsection{Modified Bo1}
\label{sec:modified-bo1}
Taking the Bo1 formula as a starting point, we modify it as follows to
obtain a more effective scoring function for an expansion term $t$. First,
an occurrence of $t$ in a top-ranked document is considered more important
than an occurrence in a lower ranked document. Thus, instead of using
$\mathit{tf}(t,d)$ directly, we scale the term frequencies by the
normalized similarity score of the corresponding document. We then
incorporate inverse collection frequency information as shown in
Equation~\ref{bo1new}.

While the \emph{tf} factor in Equation~\eqref{bo1new} is indicative of the
distribution of $t$ in the top ranked set, the \emph{ictf} factor reflects
the distribution of the term in the collection.   
\begin{equation}\label{bo1n}
  \mathit{ictf}(t) = \log_{10} \left(\frac{1}{p_c(t)} \right)
\end{equation}
\begin{equation} \label{bo1new}
  S(t) = \sum_{d \in \mathit{PRD}} {\Bigg(}\mathit{tf}(c,d)* \frac{\mbox{Sim}(d,Q)}{\underset{d^\prime \in PRD}{\operatorname{{max}~}}\mbox{Sim}(d^\prime,Q)}{\Bigg)} * \frac{\mathit{ictf}(t)}{1+\mathit{ictf}(t)}\\
\end{equation}
Finally, Equations~\ref{s-orig} to \ref{f-score} are used to merge the
original query with the expansion terms.

\subsubsection{Association based method I: LCA}
\label{sec:lca}
LCA~\cite{lca} is one of the most well-known association based term
selection methods. This method also considers all terms from the
top ranked set as candidate expansion terms. Equations~\ref{old-idf}
through \ref{fcq} show how the co-occurrence is calculated for a candidate
term $t$ and a query $Q$ consisting of terms $q_1, \ldots, q_k$ ($N_t$
has its obvious meaning, \emph{PRD} denotes the set of pseudo-relevant
documents, $n = |\mathit{PRD}|$, $\delta$ is set to 0.1 as suggested
in~\cite{lca}).
\begin{equation}
  idf_t = min(\log_{10}(N/N_t)/5.0, 1.0) \label{old-idf}
\end{equation}
\begin{equation}
  co(t,q_i)=\sum_{d \in \mathit{PRD}} \mathit{tf}(t,d)*\mathit{tf}(q_i,d) \label{co}
\end{equation}
\begin{equation}
  codegree(t,q_i)= \frac{\log_{10}(co(t,q_i)+1)*idf_t}{\log_{10}(n)} \label{codegree}
\end{equation}
\begin{equation}
  S(t) = \sum_{i=1}^{k} {idf_q}_i * \log_{10}(\delta+codegree(t,q_i)) \label{fcq}
\end{equation}
The $T$ terms with the highest $S(t)$ scores are selected as expansion
terms. Finally, the $j$-th ``best'' term is weighted according to 
Equation~\ref{9i}. 
\begin{equation}
 score_{exp}(t) = 1.0 - \frac{0.9*j}{T} \label{9i} \\
\end{equation}
We did not use noun-phrases or passage level retrieval, since the authors
show that these refinements do not have much impact. Our experiments
confirm that our implementation yields very similar results for the
collections and settings mentioned in \cite{lca}.

\subsubsection{Modified LCA}
\label{sec:modified-lca}
Our implementation of the above formulae did not yield the expected
improvements. A failure analysis suggests that Equation~\ref{co} might be
the culprit. For example, consider the TREC4 query: ``How has affirmative
action affected the construction industry?''. Two terms \emph{papuc}
(Pennsylvania Public Utility Commission) and \emph{limerick} are very
highly ranked among candidate terms by Equation~\ref{fcq}, even though
these are not useful expansion terms. This is because in one top document,
the word `papuc' occurs 21 times and a query word (`construction') occurs
35 times. The multiplication of raw term frequencies in Equation~\ref{co}
results in a very high weight for the term `papuc'. A similar problem
occurs in case of `limerick', which occurs 17 times in one document. 

Our hypothesis is that the number of co-occurrences of a term pair can only
be as large as the minimum term frequency of the two terms under
consideration. We also hypothesize that co-occurrences in a document are
more important if the document is ``close'' (or similar) to the query. 
Finally, we use the \emph{idf} factor\footnote{Note that we use Robertson's
  \emph{idf} formula~\cite{idf} (Equation~\ref{new-idf}) instead of
  Equation~\ref{old-idf}.} for a candidate expansion term when calculating
its co-occurrence (Equation~\ref{new-co}); it is no longer used when
calculating co-degree (Equation~\ref{new-codegree}).  
Equations~\ref{new-idf}--\ref{new-fcq} define our modified approach for
calculating the association between a candidate term $t$ and the query
$Q$.
\begin{equation}
  idf_t = \log_{10}\frac{N-N_t+0.5}{N_t+0.5} \label{new-idf} \\
\end{equation}
\begin{equation}\label{new-co}
\begin{split}
 co(t,q_i)& = \sum_{d \in \it{PRD}} {\Bigg(} \min(\it{tf}(t,d),\it{tf}(q_i,d)) *  \\
          & \max(idf_{t\vee q_i},0)  * \frac{\mbox{Sim}(d,Q)}{\underset{d^\prime \in PRD}{\operatorname{{max}~}}\mbox{Sim}(d^\prime,Q)} {\Bigg)} 
\end{split}
\end{equation}
where $idf_{t\vee q_i}$ denotes the idf of term $t$ or $q_i$, based on whose term frequency is minimum in document $d$.
\begin{equation}
  codegree(t,q_i)= \frac{\log_{10}(co(t,q_i)+1)}{\log_{10}(n)} \label{new-codegree}\\
\end{equation}
\begin{equation}
  S(t) = \sum_{i=1}^{k} {idf_q}_i * \log_{10}(\delta+codegree(t,q_i)) \label{new-fcq} \\
\end{equation}
As before, the $T$ terms with the highest association scores ($S(t)$) are
selected as expansion terms. The final term weights in the expanded query
are determined using Equations~\ref{s-orig} to \ref{f-score}.

\subsubsection{Association based method II: RM3}
\label{sec:rm3}
Relevance-based language models~\cite{rblm,rm3} constitute a more recent
association-based approach. In this approach, the association $S(t)$ between
a word $t$ and a query $Q = q_1, \ldots, q_k$ can be measured by $P(t, q_1,
\ldots, q_k)$, the joint probability of observing the word together with
the query words, when these words are all sampled from an (unknown)
relevance model. This relevance model consists of a finite universe
$\cal{M}$ of unigram distributions each of which corresponds to a (pseudo)
relevant document. Under the assumption that $t, q_1, \ldots, q_k$ are
independently and identically sampled from $M\in\cal{M}$,
\begin{eqnarray} \label{rm3-equ}
  S(t) & = & P(t,q_1,\ldots,q_k) \nonumber \\
       & = & \sum_{M \in \cal{M}}P(M)P(t|M)\prod_{i=1}^{k}P(q_i|M) \nonumber \\
       & = & \frac{1}{\#PRD}\sum_{d \in \it{PRD}} 
             {\Bigg(}\frac{\mathit{tf}(t,d)}{|d|} \times 
             \prod_{i=1}^{k} \frac{\mathit{tf}(q_i,d)+ \mu P(q_i|C)}{|d|+\mu}
             {\Bigg)}, 
\end{eqnarray}
where $\mu = 2500$ is a smoothing parameter, and $P(q_i|C) = p_c(q_i)$.
Equations~\ref{x} to \ref{z} 
show, as before, how the expanded query terms are added to the original
query. This implementation duplicates the LEMUR RM3 method. However, we
used the i.i.d.\ sampling approach instead of the conditional sampling
method recommended in~\cite{rblm}, since this gave us better results.
\begin{equation} \label{x}
  score_{exp}(t) = \frac{S(t)}{\sum\limits_{d \in {PRD}}\sum\limits_{t^\prime \in d} S(t^\prime)}
\end{equation}
\begin{equation} \label{y}
  score_{orig}(t) = \frac{\mathit{tf}(t,Q)}{|Q|}
\end{equation}
\begin{equation} \label{z}
  score(t) = \alpha * score_{exp}(t) + (1 - \alpha) * score_{orig}(t),
  \mbox{where} \  0 \le \alpha \le 1
\end{equation}

\subsection{Combining association based method with distribution based method}
\label{sec:mixed} 
Section~\ref{sec:basic-methods} reviews two different types of query
expansion methods. In this section, we describe a hybrid approach that
combines the above methods to improve retrieval effectiveness.

We conducted some preliminary experiments to explore various ways to
combine individual methods. Our first attempt involved simply adding up the
normalized weights of the expansion terms as computed by the individual
methods. This particular method did not perform better than the individual
methods. Next, we tried to apply the methods sequentially: the original
query is expanded using one of the methods, and the expanded query is then
used as the initial query for the other method and expanded further. This
approach also results in a performance drop. The final approach that we
tried also applies the methods sequentially, but in a different way. One of
the methods is used first to create a large expanded query. This query is
then \emph{refined} (instead of being expanded further) using the other
method. This method turns out to work well, and yields significant
improvements over the individual methods.

We can see from Table ~\ref{kldlca-improvements} that the
distribution-based methods generally perform better than the
association-based methods on most of the test collections used in our
experiments. We therefore choose a distribution-based method --- KLD
(Equation~\ref{kld-eq}) or Bo1 (Equation~\ref{bo1new}) --- to first select
(and weight) a relatively large number of candidate terms that occur
preferentially in a few top-ranked documents, where the proportion of
relevant documents is expected to be high. This set is then refined using
co-occurrence information: terms that do not co-occur significantly with
original query terms are discarded. Conversely, candidate expansion terms
that are relatively poorly ranked by the distribution-based method have a
chance to be included in the final query if they adequately co-occur with
the original query terms. More precisely, the candidate terms are re-ranked
using an association-based method --- our modified version of LCA
(Equation~\ref{new-fcq}) or RM3 (Equation \ref{rm3-equ}) --- that looks at
a larger number of top-ranked documents. The top $T$ terms from this
re-ranked list are chosen as the final expansion terms. However, we retain
the weights of these terms as determined by the distribution based method.
As before, the final term weights in the expanded query are determined
using Equations~\ref{s-orig} to \ref{f-score}.

\section{Experimental Setup}
\label{sec:exp-set}
Table \ref{query-stats} lists the details of the test collections used in
our experiments. As real-life queries are very short, we used only the
title field of all these queries, except for the TREC4 queries, which
contain only the description field. Many of the queries thus contain only
one term, and most of the remainder are no longer than three words; only
the TREC4 queries are longer.

\begin{table}
\scriptsize
\begin{center}
\caption{Test collections \label{query-stats}}
\ \\
{
\begin{tabular}{|c|c|c|} \hline
  Query Id. & \# of Queries & Documents\\\hline
  TREC123 & 150 & TREC disks 1, 2\\
  51--200 & &\\\hline
  TREC4 & 49 & TREC disks 2, 3\\
  202--250 & &\\\hline 
  TREC5 & 50 & TREC disks 2, 4\\
  251--300 & &\\\hline
  TREC678 & 150 & TREC disks 4, 5 - CR\\
  301--450 &&\\\hline
  ROBnew & 100  & TREC disks 4, 5 - CR\\
  601--700 &&\\\hline
  TREC910 & 100  & WT10G\\
  451--550 &&\\\hline
\end{tabular}
}
\end{center}
\end{table}

We used the TERRIER\footnote{http://terrier.org/} retrieval system for our
experiments. At the time of indexing, stopwords are removed and Porter's
stemmer is used as preprocessing. All documents and queries are indexed
using single terms, no phrases are used. The IFB2 variant of the Divergence
From Randomness model~\cite{dfr} --- a relatively recent model that
performs well across test collections --- is used for term-weighting in all
our experiments as it performs better compared to the other variants available
within TERRIER. Parameters are set to the default values used in TERRIER.

Results are evaluated using standard evaluation metrics (Mean Average
Precision (MAP), precision at top 10 ranks (P@10), and overall recall
(number of relevant documents retrieved)). Additionally, for each expansion
method, we report the percentage of queries for which the method resulted
in an improvement in MAP of more than 5\% over the baseline (no feedback).

\section{Experimental Results}
\label{sec:exp-res}
We now present experimental results for the QE methods described in
Section~\ref{sec:method}. The first set of results presented in
Section~\ref{subsec:mod-meth} pertain to our implementation of well-known
QE methods, as well as the proposed refinements to these methods.
Section~\ref{subsec:com-meth} corresponds to the combination-based method
described in Section~\ref{sec:mixed}.

\parab{Notation.} We use the following labels to denote various techniques
in tables~/ figures.
\begin{table}[!h]
\scriptsize
  \centering
  \begin{tabular}{lll}\hline
    \textbf{Name} & \textbf{Description} & \textbf{Details in}\\\hline
    KLD & Our implementation of the KLD method & Section~\ref{sec:kld} \\
    Bo1 & Our implementation of the Bo1 method & Section~\ref{sec:bo1} \\
    Bo1new & Modified Bo1 method               & Section~\ref{sec:modified-bo1} \\
    LCA & Our implementation of the LCA method & Section~\ref{sec:lca} \\
    LCAnew & Modified LCA method               & Section~\ref{sec:modified-lca} \\
    RM3 & Our implementation of the RM3 method & Section~\ref{sec:rm3} \\
    KLDLCA & Combination of KLD and LCAnew & \rdelim\}{4}{1in}[Section~\ref{sec:mixed}] \\
    KLDRM3 & Combination of KLD and RM3 & \\
    Bo1LCA & Combination of Bo1new and LCAnew & \\
    Bo1RM3 & Combination of Bo1new and RM3 & \\\hline
  \end{tabular}
  \caption{Labels for various QE methods}
  \label{tab:notation}
\end{table}
In the following tables, results that are statistically significantly
better (as determined by a two-tailed paired $t$-test with a confidence
level of 95\%) than the baseline (no feedback), KLD, Bo1new, LCAnew and RM3 are marked
with the superscripts B, k, b, l and r respectively. 

\renewcommand{\parab}[1] {\bigskip \noindent \emph{#1}~}

\subsection{Experiment 1: modified methods}
\label{subsec:mod-meth}
\begin{figure}
\scriptsize
  \centering
  \epsfig{file=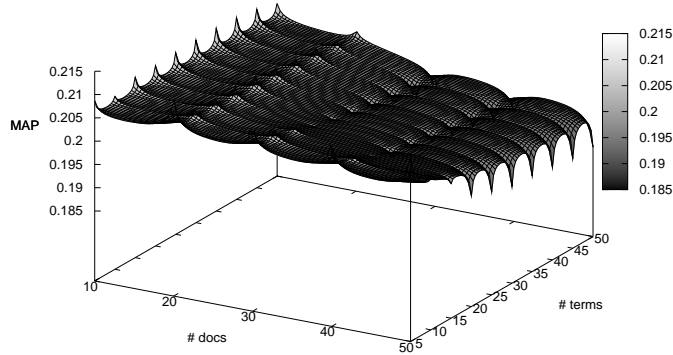, width=2.5in, angle=270}
  \caption{{\scriptsize Performance (MAP) of LCA on TREC678 for different parameter
    settings.}} 
  \label{lca-img}
\end{figure}

\parab{Baselines.} For comparison, we use the following baselines.
\begin{enumerate}
\item No feedback. The original, unexpanded queries are used for retrieval
  using the baseline method described in Section~\ref{sec:exp-set}. 
\item Bo1. For this method, Amati~\cite{amati-thesis} suggested adding
  $T=10$ expansion terms from the top $D=3$ documents. We use $T=40$ and
  $D=10$ instead, since we wanted a larger number of candidate terms,
  particularly for the combination-based method.
  Our experiments confirm that we get comparable results with these
  parameter settings.
\item LCA. To determine the parameters for LCA, we used the TREC678
  collection as a ``tuning'' dataset, as TREC678 is comparatively recent,
  and contains a large set of queries. We varied the number of top-ranked
  documents ($D$) from 10 to 50 in steps of 10, while the number of
  expansion terms ($T$) was varied from 5 to 50 in steps of 5. Xu and
  Croft~\cite{lca} recommended using $D=70$ and $T=70$. In our setup,
  however, a setting of $D=10$ documents and $T=40$ expansion terms works
  well. Figure~\ref{lca-img} shows that these settings work well in terms
  of MAP. We use these values on all collections used in our experiments. A
  similar exercise suggests that the same settings can be used for LCAnew
  as well.
\end{enumerate}

\parab{LCAnew.} Our first goal is to verify that the proposed modifications
to the LCA formula actually yield improvements in retrieval performance.
Table~\ref{lca-improvements} shows that, with `title only' queries, our
implementation of the original LCA formula results in a drop in MAP for
almost all collections. Only for the TREC4 collection (in which queries
consist of a description only), a marginal improvement is observed,
suggesting that the original method works better for longer queries.
Indeed, the experiments by Xu and Croft all used relatively long queries,
e.g., the full TREC3 queries (including title, description and narrative
fields), the TREC4 queries, and the description field of TREC5 queries.

Compared to the original formula, the modified formula results in
significant improvements in MAP across all data sets. On the ROBnew
collection in particular, it performs very well, outperforming the original
method by nearly 24\%. For the TREC4 corpus, an improvement of about
11.41\% (over LCA) is observed. The modifications thus seem to be effective
for both short as well as relatively longer queries. The LCAnew method is
also better in terms of P@10, number of relevant documents retrieved, and
robustness.

\parab{Bo1new.} Table~\ref{lca-improvements} shows that Bo1new gives better
results than Bo1 across all test collections. For the TREC123, ROBnew, and
TREC910 collections, these improvements are significant. The modified
method also yields better P@10, and appears to be more robust across all
datasets. With regard to the number of relevant documents retrieved, Bo1new
is better on most collections. Overall, Bo1new appears to be a superior
alternative to Bo1 in all respects.

\bigskip
Thus, based on Table~\ref{lca-improvements}, we conclude that LCAnew and
Bo1new are more effective, and can be used in place of LCA and Bo1. 

\begin{table*}
\scriptsize
  \begin{center}
    \begin{tabular}{| c | c | c || c | c || c | c |}
      \hline
      Dataset & Measure & Baseline & LCA & LCAnew & Bo1 & Bo1new \\ \hline

TREC123 & MAP             & 0.218 & 0.213   & 0.254$^{B*}$   & 0.272   & 0.277$^{B*}$  \\ 
        &                 &       & (-2.4)  & (16.4)  & (24.4)  & (26.6) \\ 
        & P@10            & 0.481 & 0.472   & 0.520   & 0.531   & 0.545  \\ 
        &                 &       & (-1.8)  & (8.2)   & (10.4)  & (13.4) \\ 
        & \#rel\_ret      & 16536 & 15714   & 17475   & 18227   & 18242  \\ 
        &                 &       & (-5.0)  & (5.7)   & (10.2)  & (10.3) \\ 
        & $>$ baseline on & 0     & 35      & 54      & 58      & 62     \\ \hline
TREC4   & MAP             & 0.217 & 0.219   & 0.244$^{B}$   & 0.256   & 0.259$^{B}$  \\ 
        &                 &       & (1.1)   & (12.6)  & (17.8)  & (19.5) \\ 
        & P@10            & 0.461 & 0.400   & 0.496   & 0.441   & 0.467  \\ 
        &                 &       & (-13.3) & (7.5)   & (-4.4)  & (1.3)  \\ 
        & \#rel\_ret      & 3482  & 3507    & 3691    & 3854    & 3768   \\ 
        &                 &       & (0.7)   & (6.0)   & (10.7)  & (8.2)  \\ 
        & $>$ baseline on & 0     & 38      & 57      & 55      & 57     \\ \hline
TREC5   & MAP             & 0.157 & 0.130   & 0.152$^{*}$   & 0.166   & 0.168  \\ 
        &                 &       & (-17.6) & (-3.1)  & (5.4)   & (7.0)  \\ 
        & P@10            & 0.286 & 0.210   & 0.238   & 0.248   & 0.270  \\ 
        &                 &       & (-26.6) & (-16.8) & (-13.3) & (-5.6) \\ 
        & \#rel\_ret      & 1936  & 1894    & 2053    & 2194    & 2183   \\ 
        &                 &       & (-2.2)  & (6.0)   & (13.3)  & (12.8) \\ 
        & $>$ baseline on & 0     & 20      & 38      & 42      & 44     \\ \hline
TREC678 & MAP             & 0.218 & 0.209   & 0.250$^{B*}$   & 0.255   & 0.257$^{B}$  \\ 
        &                 &       & (-4.2)  & (14.8)  & (16.8)  & (17.7) \\ 
        & P@10            & 0.431 & 0.379   & 0.420   & 0.427   & 0.436  \\ 
        &                 &       & (-12.2) & (-2.6)  & (-0.9)  & (1.1)  \\ 
        & \#rel\_ret      & 7287  & 7367    & 8152    & 8529    & 8463   \\ 
        &                 &       & (1.1)   & (11.9)  & (17.0)  & (16.1) \\ 
        & $>$ baseline on & 0     & 36      & 52      & 53      & 60     \\ \hline
ROBnew  & MAP             & 0.278 & 0.264   & 0.327$^{B*}$   & 0.307   & 0.331$^{B*}$  \\ 
        &                 &       & (-5.0)  & (17.6)  & (10.3)  & (19.0) \\ 
        & P@10            & 0.421 & 0.385   & 0.452   & 0.394   & 0.433  \\ 
        &                 &       & (-8.6)  & (7.2)   & (-6.5)  & (2.9)  \\ 
        & \#rel\_ret      & 2887  & 2864    & 3009    & 3178    & 3202   \\ 
        &                 &       & (-0.8)  & (4.2)   & (10.1)  & (10.9) \\ 
        & $>$ baseline on & 0     & 36      & 53      & 48      & 56     \\ \hline
TREC910 & MAP             & 0.195 & 0.155   & 0.175   & 0.189   & 0.202$^{*}$  \\ 
        &                 &       & (-20.6) & (-10.6) & (-3.3)  & (3.5)  \\ 
        & P@10            & 0.307 & 0.231   & 0.291   & 0.284   & 0.304  \\ 
        &                 &       & (-24.9) & (-5.3)  & (-7.6)  & (-1.0) \\ 
        & \#rel\_ret      & 3770  & 3440    & 3646    & 3974    & 3948   \\ 
        &                 &       & (-8.8)  & (-3.3)  & (5.4)   & (4.7)  \\ 
        & $>$ baseline on & 0     & 27      & 33      & 41      & 45     \\ \hline
    \end{tabular}
  \end{center} 
  \caption{Improvements on different datasets obtained by modifying LCA /
    Bo1. The ``$>$ baseline on'' line shows the \textbf{\%-age} of queries
    on which each method beats the baseline by $>5\%$. A * denotes an
    improvement (over original formula) that is statistically significant.}
  \label{lca-improvements}
\end{table*}

\subsection{Experiment 2: combination methods}
\label{subsec:com-meth}
As explained in Section~\ref{sec:mixed}, in the combination-based approach,
we first select a large set of candidate terms ($T=100$) from $D=10$
documents using a distribution-based QE method. The association of these
candidate terms with the query terms is computed using the top $D'=50$
documents\footnote{Measuring association scores over the top 30--50
  documents works about equally well.}, and the best $T'=40$ terms (as
determined by an association-based method) are included in the final query.
We report results for a total of $2 \times 2 = 4$ combinations: KLDLCA (LCAnew
with KLD), KLDRM3 (RM3 with KLD), Bo1LCA (LCAnew with Bo1new), and Bo1RM3 (RM3
with Bo1new).

\parab{Baselines.} We compare the combination-based methods with the
following baselines.
\begin{enumerate}
\item No feedback. Same as in Section~\ref{subsec:mod-meth}
\item KLD. We find that a setting of $D = 10$ top-ranked documents and
  $T=40$ expansion terms works well for KLD across collections. This is in
  agreement with the observations of Carpineto et al.\cite{carpineto-kld}.

  Note that the results presented here correspond to our implementation of
  KLD within TERRIER. While our implementation provides better results than
  TERRIER's native implementation of KLD, we were not able to exactly
  replicate the results reported in~\cite{carpineto-kld}. This is likely
  due to differences between the retrieval functions, indexing or query
  processing. For example, using full queries (title, desc and narr) on the
  TREC8 collection, and BM25 as the base term-weighting formula, we get MAP
  scores of 0.2992 for KLD (compared to a baseline of 0.2625). When using
  the IFB2 model, however, the baseline is higher (MAP = 0.2753), but KLD
  appears less effective (MAP = 0.2850).
\item Bo1new, LCAnew. As discussed in Section~\ref{subsec:mod-meth}, for
  these methods also, we use $D = 10$ documents, and $T = 40$ terms. 
\item RM3. We use $D = 50$ documents (as suggested in~\cite{rblm}) and 
  $T = 50$ terms . We set the Dirichlet smoothing parameter ($\mu$) to 2500
  and the interpolation parameter to 0.5, based on the default settings for
  these parameters in Lemur~\footnote{http://www.lemurproject.org/}. As before, we used the TREC678
  collection to verify that these parameter values work well for us. In
  fact, for a number of datasets, our results for RM3 are superior to those
  reported in other recent papers (\cite{RM3-check}, for example).
\end{enumerate}

\begin{sidewaystable}
\scriptsize
  \begin{center}
    \begin{tabular}{|c|c|c|c|c|c|c|c|c|c|c|} \hline

Dataset & Measure    & Baseline & KLD    & Bo1new & LCAnew  & RM3    & KLDLCA & KLDRM3 & Bo1LCA & Bo1RM3 \\ \hline
TREC123 & MAP        & 0.218    & 0.274  & 0.277  & 0.254   & 0.249  & 0.280$^{Blr}$  & 0.277$^{Blr}$  & 0.285$^{Bklr}$  & 0.284$^{Blr}$  \\ 
        &            &          & (25.4) & (26.6) & (16.4)  & (14.1) & (28.0) & (26.8) & (30.6) & (29.8) \\ 
        & P@10       & 0.481    & 0.537  & 0.545  & 0.520   & 0.511  & 0.537  & 0.541  & 0.553  & 0.540  \\ 
        &            &          & (11.8) & (13.4) & (8.2)   & (6.2)  & (11.8) & (12.5) & (15.0) & (12.3) \\ 
        & \#rel\_ret & 16536    & 18299  & 18242  & 17475   & 17702  & 18585  & 18438  & 18701  & 18639  \\ 
        &            &          & (10.7) & (10.3) & (5.7)   & (7.1)  & (12.4) & (11.5) & (13.1) & (12.7) \\ 
        & $>$ baseline on  & 0        & 62     & 62     & 54      & 64     & 67     & 65     & 67     & 68     \\ 
\hline
TREC4   & MAP        & 0.217    & 0.261  & 0.259  & 0.244   & 0.252  & 0.279$^{Bkblr}$  & 0.265$^{B}$  & 0.273$^{Bl}$  & 0.265$^{B}$  \\ 
        &            &          & (20.2) & (19.5) & (12.6)  & (15.9) & (28.7) & (22.3) & (25.6) & (21.9) \\ 
        & P@10       & 0.461    & 0.455  & 0.467  & 0.496   & 0.516  & 0.498  & 0.480  & 0.498  & 0.502  \\ 
        &            &          & (-1.3) & (1.3)  & (7.5)   & (11.9) & (8.0)  & (4.0)  & (8.0)  & (8.8)  \\ 
        & \#rel\_ret & 3482     & 3815   & 3768   & 3691    & 3689   & 3882   & 3781   & 3846   & 3775   \\ 
        &            &          & (9.6)  & (8.2)  & (6.0)   & (5.9)  & (11.5) & (8.6)  & (10.5) & (8.4)  \\ 
        & $>$ baseline on  & 0        & 57     & 57     & 57      & 75     & 55     & 59     & 57     & 61     \\ 
\hline
TREC5   & MAP        & 0.157    & 0.168  & 0.168  & 0.152   & 0.170  & 0.171  & 0.172$^{k}$  & 0.174$^{l}$  & 0.173  \\ 
        &            &          & (6.9)  & (7.0)  & (-3.1)  & (8.2)  & (9.0)  & (9.2)  & (10.4) & (9.9)  \\ 
        & P@10       & 0.286    & 0.268  & 0.270  & 0.238   & 0.336  & 0.274  & 0.280  & 0.290  & 0.304  \\ 
        &            &          & (-6.3) & (-5.6) & (-16.8) & (17.5) & (-4.2) & (-2.1) & (1.4)  & (6.3)  \\ 
        & \#rel\_ret & 1936     & 2184   & 2183   & 2053    & 2077   & 2218   & 2166   & 2226   & 2184   \\ 
        &            &          & (12.8) & (12.8) & (6.0)   & (7.3)  & (14.6) & (11.9) & (15.0) & (12.8) \\ 
        & $>$ baseline on  & 0        & 42     & 44     & 38      & 50     & 52     & 50     & 48     & 48     \\ 
\hline
TREC678 & MAP        & 0.218    & 0.257  & 0.257  & 0.250   & 0.230  & 0.266$^{Bkblr}$  & 0.260$^{Br}$  & 0.265$^{Bblr}$  & 0.259$^{Br}$  \\ 
        &            &          & (18.0) & (17.7) & (14.8)  & (5.6)  & (22.0) & (19.2) & (21.6) & (18.7) \\ 
        & P@10       & 0.431    & 0.438  & 0.436  & 0.420   & 0.435  & 0.441  & 0.431  & 0.435  & 0.428  \\ 
        &            &          & (1.6)  & (1.1)  & (-2.6)  & (0.8)  & (2.2)  & (0.0)  & (0.8)  & (-0.8) \\ 
        & \#rel\_ret & 7287     & 8556   & 8463   & 8152    & 7617   & 8567   & 8552   & 8570   & 8449   \\ 
        &            &          & (17.4) & (16.1) & (11.9)  & (4.5)  & (17.6) & (17.4) & (17.6) & (15.9) \\ 
        & $>$ baseline on  & 0        & 52     & 60     & 52      & 45     & 57     & 57     & 61     & 58     \\ 
\hline
ROBnew  & MAP        & 0.278    & 0.312  & 0.331  & 0.327   & 0.305  & 0.326$^{Bkr}$  & 0.322$^{Bk}$  & 0.341$^{Bkbr}$  & 0.341$^{Bkbr}$  \\ 
        &            &          & (12.2) & (19.0) & (17.6)  & (9.8)  & (17.2) & (15.9) & (22.5) & (22.6) \\ 
        & P@10       & 0.421    & 0.405  & 0.433  & 0.452   & 0.442  & 0.438  & 0.424  & 0.455  & 0.455  \\ 
        &            &          & (-3.8) & (2.9)  & (7.2)   & (5.0)  & (4.1)  & (0.7)  & (7.9)  & (7.9)  \\ 
        & \#rel\_ret & 2887     & 3172   & 3202   & 3009    & 3002   & 3173   & 3160   & 3214   & 3218   \\ 
        &            &          & (9.9)  & (10.9) & (4.2)   & (4.0)  & (9.9)  & (9.5)  & (11.3) & (11.5) \\ 
        & $>$ baseline on  & 0        & 52     & 56     & 53      & 56     & 55     & 57     & 62     & 63     \\ 
\hline
TREC910 & MAP        & 0.195    & 0.193  & 0.202  & 0.175   & 0.211  & 0.204$^{kl}$  & 0.210$^{kl}$  & 0.207$^{kl}$  & 0.213$^{kl}$  \\ 
        &            &          & (-1.1) & (3.5)  & (-10.6) & (8.0)  & (4.7)  & (7.4)  & (6.0)  & (9.1)  \\ 
        & P@10       & 0.307    & 0.293  & 0.304  & 0.291   & 0.329  & 0.313  & 0.309  & 0.320  & 0.313  \\ 
        &            &          & (-4.6) & (-1.0) & (-5.3)  & (7.0)  & (2.0)  & (0.7)  & (4.3)  & (2.0)  \\ 
        & \#rel\_ret & 3770     & 3987   & 3948   & 3646    & 3889   & 4021   & 3992   & 4016   & 4018   \\ 
        &            &          & (5.8)  & (4.7)  & (-3.3)  & (3.2)  & (6.7)  & (5.9)  & (6.5)  & (6.6)  \\ 
        & $>$ baseline on  & 0        & 44     & 45     & 33      & 53     & 51     & 50     & 53     & 48     \\ \hline

    \end{tabular}
  \end{center} 
  \caption{Improvements on different datasets obtained by combining
    association based and distribution based QE methods.
    (The ``$>$ baseline on'' line shows the \textbf{\%-age} of queries on
    which each method beats the baseline by $>5\%$.)}
  \label{kldlca-improvements}
\end{sidewaystable}

Table~\ref{kldlca-improvements} shows that the proposed combined approaches
are statistically significantly better than the no-feedback method across
all test collections except for TREC5 and TREC910. More importantly, the
combined methods consistently work better than the individual QE methods
involved in the combination, as well as most of the other standard QE
methods. These differences are, by and large, statistically significant,
with only a few exceptions. Overall, while RM3 seems to be the best in
terms of P@10 in most of the cases, the combination based methods are
generally the best on all other measures. We now briefly discuss each
combination in turn.

\parab{KLDLCA.} KLDLCA is better than KLD or LCAnew alone on all measures, and
across all datasets. For 5 out of the 6 collections, the combination yields
significant improvements in MAP over KLD or LCAnew or both. It is interesting
to note that for the sixth collection (TREC5), LCAnew results in a drop in
performance compared to the no-expansion baseline. However, the
combinations KLDLCA and Bo1LCA perform better than the baseline as well as
KLD.

In general, the combination also seems to be \emph{safer}, in the sense
that combination-based expansion usually hurts fewer queries than expansion
using either KLD or LCAnew. 
On a related note, a query wise analysis of the TREC678 dataset shows that
out of the 150 queries in this collection, there are 59 queries on which
KLD outperforms KLDLCA (with an average improvement in MAP of 0.0148), but
KLDLCA does better than KLD on 85 queries, and improves MAP by 0.0255 on
average. Similarly, LCAnew performs better than KLDLCA on 68 queries (average
improvement in MAP = 0.0360), whereas KLDLCA wins on 81 queries (average
improvement in MAP = 0.0594).

It is particularly encouraging that KLDLCA is also generally better than
the two other state-of-the-art QE methods, RM3 and Bo1new, on all measures and
across all datasets. The only exceptions are: RM3 yields better P@10 on
TREC4,TREC5, ROBnew and TREC910
and superior MAP for TREC910, while Bo1 outperforms KLDLCA on P@10 for
TREC123, on MAP for ROBnew, and on the number of relevant documents retrieved
for ROBnew.

\parab{KLDRM3.} KLDRM3 also yields better MAP than either KLD or RM3 on all
collections (but neither difference is statistically significant for
TREC4). It is also better than the other individual QE methods 
(LCAnew and Bo1new) on all corpora except ROBnew, where Bo1new outperform KLDRM3.
This method is among the safest: only Bo1new yields improvements on marginally
more queries for the TREC678 collection; on all other datasets, expansion
by KLDRM3 improves performance on more queries than any other method.

\parab{Bo1LCA, Bo1RM3.} Both methods yield improvements (often significant)
in MAP compared to all individual QE methods. Indeed, with a few
exceptions, Bo1LCA is better than all individual QE methods for all the
datasets and on all the measures.

\subsection{Discussion}
The results in the preceding section confirm our hypothesis that, on
average, distribution and association based methods work well together. For
queries such as 321 (\emph{Women in Parliaments}), the combination works as
expected. Both LCAnew (AP = 0.2531) and KLD (AP = 0.2611) select and
assign relatively high weights to specific names such as \emph{mashokw},
\emph{jankowska}, \emph{starkova}, \emph{fedulova}. When LCAnew is used to
filter terms based on association information obtained from 50 documents,
these terms are eliminated, and retrieval effectiveness goes up (AP = 0.3629).

More interesting are queries where the combination fails. Query 350
(\emph{Health and Computer Terminals}) is one such example for which 
LCAnew (AP = 0.5911) and KLD (AP = 0.4512) both do reasonably well, but AP
drops to 0.4007 for KLDLCA. For this particular query, filtering candidates
terms using association information results in the elimination of a number
of good expansion terms. 

Unfortunately, no general pattern seems to be discernible for such queries
where a combination is inferior to either or both of its ingredients.

\section{Conclusion}
\label{sec:conclusion}
In this study, our objective was to combine distribution based and
association based query expansion methods. Using a number of standard test
collections, we have shown that distribution based QE can be improved by
using an association based method to refine term selection. The proposed
combination gives better results than each individual method, as well as
other state-of-the-art approaches.

En route to this goal, we also proposed some modifications to a few
well-known QE methods which lead to improved performance. This may be
regarded as an additional contribution of this paper.

In future work, we intend to do a more comprehensive study by investigating
other combinations of QE methods.

\bibliographystyle{plain}
\bibliography{qe-pal}

\end{document}